# Towards Proxemic Mobile Collocated Interactions[1]

Andrés Lucero, Aalto University, Helsinki, Finland
Marcos Serrano, University of Toulouse - IRIT, Toulouse, France

**ABSTRACT**

Research on mobile collocated interactions has been exploring situations where collocated users engage in collaborative activities using their personal mobile devices (e.g., smartphones and tablets), thus going from personal/individual toward shared/multiuser experiences and interactions. The proliferation of ever-smaller computers that can be worn on our wrists (e.g., Apple Watch) and other parts of the body (e.g., Google Glass), have expanded the possibilities and increased the complexity of interaction in what we term "mobile collocated" situations. Research on F-formations (or facing formations) has been conducted in traditional settings (e.g., home, office, parties) where the context and the presence of physical elements (e.g., furniture) can strongly influence the way people socially interact with each other. While we may be aware of how people arrange themselves spatially and interact with each other at a dinner table, in a classroom, or at a waiting room in a hospital, there are other less-structured, dynamic, and larger-scale spaces that present different types of challenges and opportunities for technology to enrich how people experience these (semi-) public spaces. In this article, the authors explore proxemic mobile collocated interactions by looking at F-formations in the wild. They discuss recent efforts to observe how people socially interact in dynamic, unstructured, non-traditional settings. The authors also report the results of exploratory F-formation observations conducted in the wild (i.e., tourist attraction).

## MOBILE COLLOCATED INTERACTIONS

Mobile devices such as smartphones and tablets were originally conceived and have traditionally been utilized for individual use. Research on mobile collocated interactions (Lucero et al., 2013; Lucero et al., 2016a) has been exploring situations in which collocated users engage in collaborative activities using their mobile devices, thus going from personal/individual multi-device workflows (Santosa & Wigdor, 2013) toward shared/multiuser experiences and interactions.

Early research on mobile collocated interactions often encouraged people to share their devices to create a collective experience or reach a common goal. Various physical and social contexts of use were taken into account, such as teamwork at the office (Lucero et al., 2010), sharing media content at home and outdoors (Clawson et al., 2008), and public expression in a theme park (Durrant et al., 2011) and in a pub (Lucero et al., 2013). More recently, researchers have been looking into simple ways to bind devices together (Jokela et al., 2015), and have conducted ethnographic work to understand the use of various mobile devices in collocated interactions (Porcheron et al., 2016). Most of this first-wave research initially looked at the use of smartphones (and tablets) to study mobile collocated interactions, and thus tended to be device-centric (Lucero et al., 2016a).

The proliferation of ever-smaller computers that can be worn on our wrists (e.g., Apple Watch) and other parts of the body (e.g., Google Glass, Microsoft HoloLens), have expanded the possibilities and increased the complexity of interaction in what we term "mobile collocated" situations. These include novel gestural interactions with wearables (Perrault et al., 2013) and interactions distributed between wearables and handheld devices (Houben and Marquardt, 2015). As wearables gain popularity, contexts in which groups of people are wearing and interacting with multiple wearable devices on their body are becoming more commonplace. In those situations, people can use a rich ecosystem (Terrenghi et al., 2009) of wearables that support collaborative tasks and experiences through multi-user applications. Such novel mobile collocated interactions may include clothing, accessories, prosthetics, and jewelry. One such example is It's About Time (Pearson et al., 2015), which explores extending smartwatch interactions to turn personal wearables into public displays. This current second-wave of mobile collocated interactions is experience-centric (Lucero et al., 2016a).

A third wave of mobile collocated interactions research should address the pressing need to understand the importance of spatial relationships between people and the digital devices in space (Lucero et al., 2016a). Adopting ideas of proxemics could allow designers to better shape each individual's personal motivations and perceptions of their interactions with both devices and others, to better support their experiences.

## PROXEMIC INTERACTIONS

Proxemics, as defined by anthropologist Edward Hall, is a research area focused on the culturally dependent use of space and physical measures (e.g., distance, orientation, and posture) to mediate and comprehend interpersonal interactions (Hall, 1963). The knowledge of proxemics has long been employed in other disciplines such as architecture, although its use in HCI is a relatively recent addition (e.g., Greenberg et al., 2014; Kortuem et al., 2005; Mueller, et al., 2014). One particularly pertinent aspect of the theory is that of proxemic 'zones', which are essentially boundaries of people's interpretations of interpersonal distance defined as intimate (less than 1.5 feet), personal (1.5– 4 feet), social (4–12 feet), and public (12–25 feet).

Proxemics prototypes have been developed that exploit knowledge of the configuration of devices and people (F-formations) in personal and group settings. The term F-formation (or facing formation) was coined by Kendon (1990) to describe the spatial arrangement of people in social encounters. Early work on F-formations focused on interpersonal interactions around large public displays and/or physical structures. Examples of this include the work by Marquardt et al. (2012) who observed groups of participants performing joint activities around a tabletop interface in a tourist information center, or the work by Paay et al. (2015), who studied F-formations in kitchens, focusing on the architectural design of the kitchen. This first wave of studies revealed that F-formations vary with the task and that physical structures in the space encourage certain formations (Marshall et al., 2011a; Marshall et al., 2011b).

Greenberg et al. (2011), in highlighting the importance of adopting proxemics to help realize the UbiComp vision of technologies that are indistinguishable from everyday life, state that *"[people] naturally expect increasing connectivity and interaction possibilities as they bring their devices in close proximity to one another."* This vision drives the idea that as we move through space, the ways in which we understand and interact with our devices should change also, essentially adopting Hall's idea of proxemic zones.

## PROXEMIC MOBILE COLLOCATED INTERACTIONS

As HCI moves towards embracing and actualizing the ideas of proxemics, we are also motivated by the idea of proxemics being used to support mobile collocated interactions, to allow our devices to not only react to presence and interaction, but also other indicators, such as the interpersonal distance people naturally use in their everyday interactions. We are also interested in studying these proxemic mobile collocated interactions in the wild.

Recent work has observed F-formations in the wild (Tong et al. 2016) in the context of an orienteering mobile learning game. They observed similar F-formations than previous works, but noticed that in a mobile environment these formations were highly dynamic, changing over time. This work unveils the importance of focusing on transitions between arrangements in mobile context and on the importance of moving from controlled observations to studies in the wild. There is limited knowledge on what could be learned about F-formations in the real world during everyday uncontrolled activities. This information could be of interest for the design of mobile collocated interactions exploiting F-formations.

## EXPLORATORY OBSERVATIONS

In this section we report on informal observations made during a workshop on *Interaction Techniques for Mobile Collocation* (Lucero et al., 2016b) at MobileHCI 2016 on September 6, 2016 in Florence, Italy. Inspired by the work presented at the workshop on dynamic F-formations in non-traditional settings (e.g., Serna et al., 2016), the workshops organizers (i.e., the authors) together with the workshop participants decided to go out and make exploratory F-formation observations in the wild. Participants (n=12) were to observe and make annotations of anything that would seem unusual or that had not been previously reported in the aforementioned studies of F-formations in controlled settings, some of which were discussed during the workshop. Such observations could include information on group sizes, how groups move in an open space, physical distance between people, or their potential use of devices.

### Procedure

After briefly scouting for a few places to go (e.g., central railway station, *Ponte Vecchio*), participants were split into three groups of four and were asked to observe formations of tourists around the Dome of Florence Cathedral (i.e., *Il Duomo*) pedestrian area. As it was high season, the weather was mostly sunny and the temperatures reached 33 degrees Celsius, the place was swarmed by tourists (Figure 1). This both provided plenty of opportunities to make observations (Figure 2), and allowed participants to mingle with the crowd. The groups split and went their separate ways to make observations for about 45 minutes. Observations were documented in form of sketches, annotations, photos, and video. The groups got back together and shared their insights with the rest. We report the findings from two of these groups.

### Results

As was mentioned earlier, the place was extremely busy. We observed a mix of tourists and locals doing different activities in this space: tourists walking alone or as small groups, shoppers carrying bags, persons resting by sitting on or lying down on benches, families carrying suitcases or with strollers, people riding or walking next to their bikes, persons walking dogs, mobile street artists selling their work. Some of these activities were performed individually, while others were done in different group sizes.

First, we discuss activities that are closely related to tourism and thus one would expect to encounter in such a context. Perhaps the often-most encountered F-formation was people gathered in a semi-circular or less orderly layered formation around a tourist guide (Figure 3) to listen to what the guide had to say about the area. The size of these groups varied greatly (i.e., between 2 and 15 persons) and the guide often used an umbrella, a flag, or other means to keep people together as they moved from one location to another.

**Figure 1. Il Duomo area in Florence, Italy swarmed by tourists where observations were made**

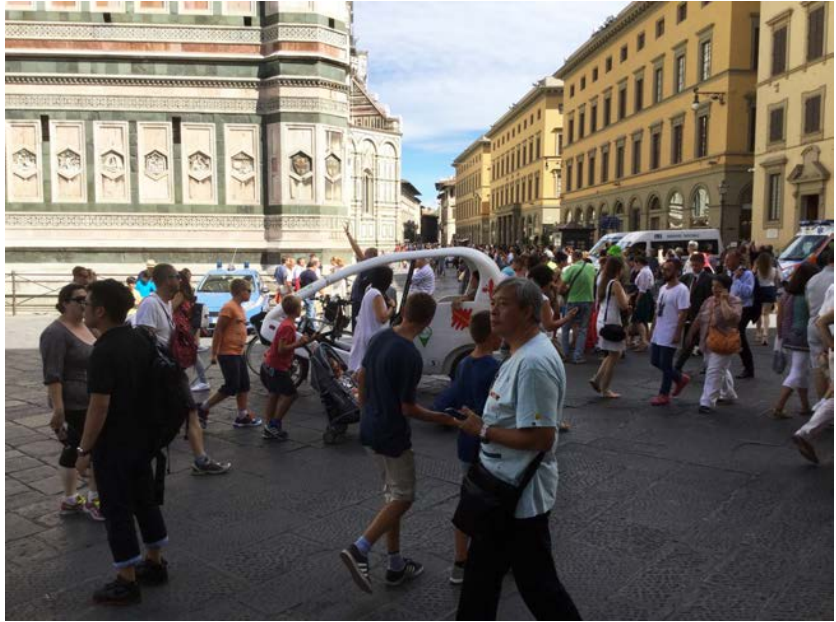

**Figure 2. A wide variety of F-formation types could be found around every corner. People asking for directions vis-à-vis, resting on a bench back-to-back, side-by-side, in an inverted I-shape and lying flat. In the background, a family creates a circular formation on the stairs.**

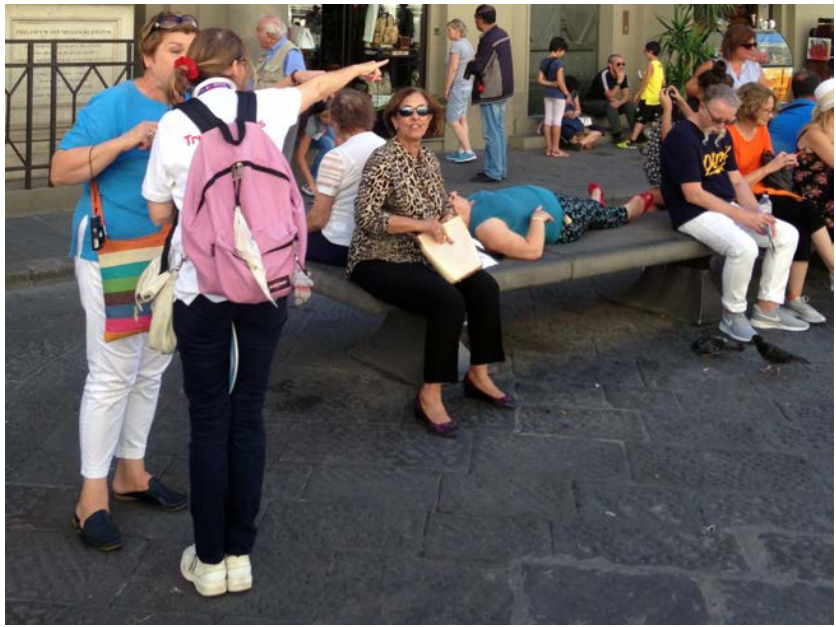

The second most common F-formations emerged while taking pictures. Here again, the size of the groups varied greatly from pairs taking selfies (Figure 4, left), to large group photos. Figure 4 (left) shows a situation where two persons are trying to take a selfie in front of the Duomo as a guided tour is passing by. Some members of the guided tour decide to walk in front of them to avoid appearing in the selfie (i.e., the lady in the pink dress), while others do not seem to notice or do not seem to care (i.e., the man pushing a stroller right behind them). This example nicely illustrates the dynamic nature of F-formations in such contexts: groups' sizes and their physical arrangement are often altered as people move from one place to another. While most photo-related situations created side-by-side formations with a single photographer in front of them, in other cases triangular F-formations were created (Figure 4, right). In this case, the two photographers are crouching to capture the Duomo behind this person, creating an F-formation where people are standing at different heights with respect to each other.

**Figure 3. People gathered in a semi-circular layered formation around a tourist guide.**

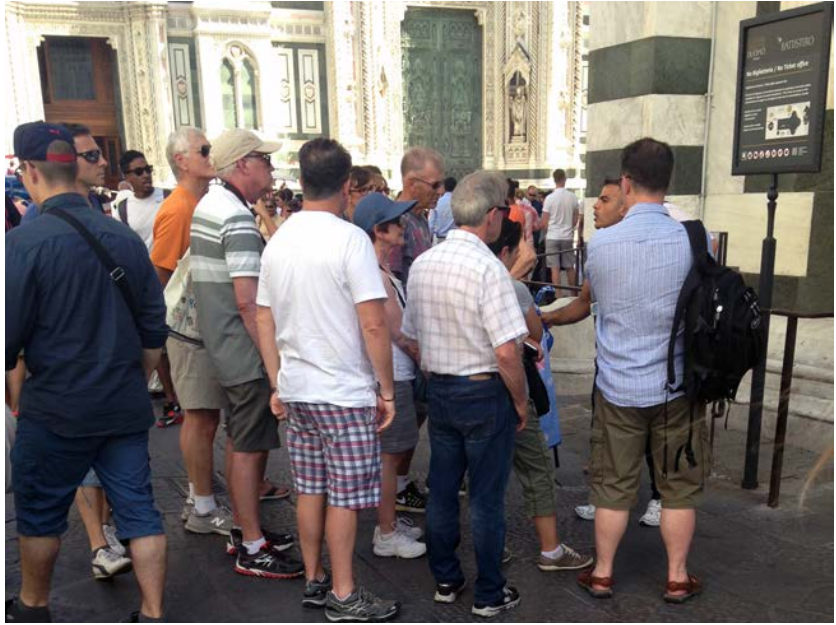

**Figure 4. People taking pictures. Two people side by side attempt to take a selfie, while people on a guided tour walk around them (left). Two people crouching to take a picture of the same person and the Duomo behind her created a triangular F-formation at different heights (right).**

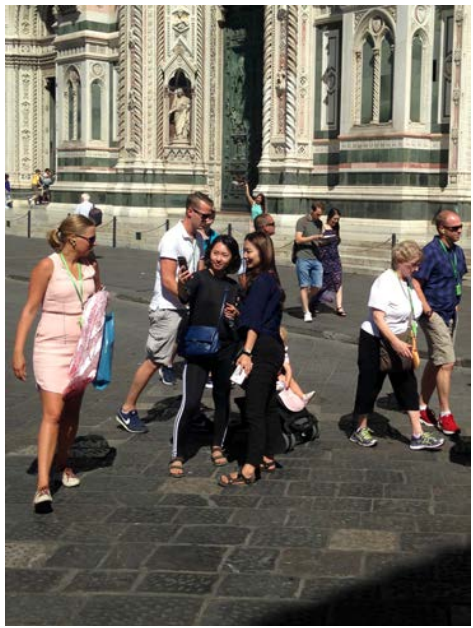
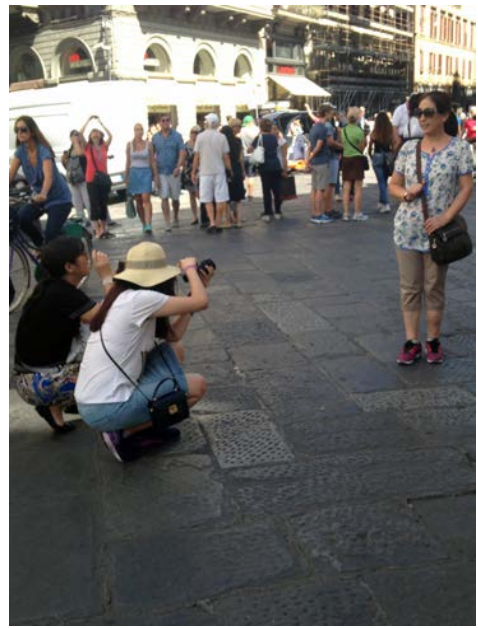

**Figure 5. Tourist guides showing directions on a map by standing vis-à-vis in front of the tourist holding the map so that the latter can see the map in the right orientation. A second tourist takes a less active role beside them, thus creating a triangular formation.**

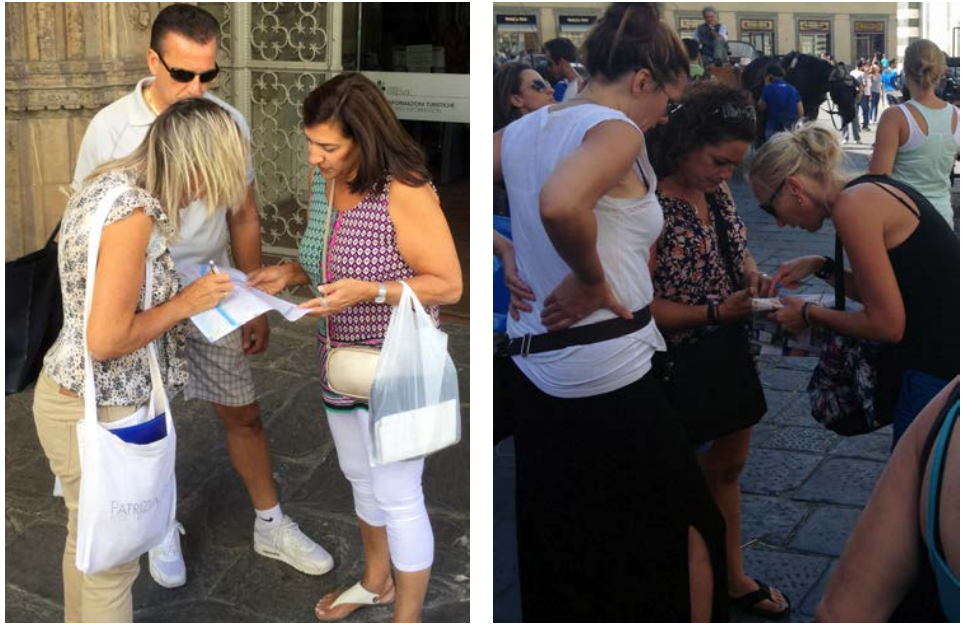

The third most common F-formation occurred when people asked for directions, often to tourist guides (Figure 5). The size of the group varied depending on how many people were asking for directions (i.e., usually one or two persons). We did notice that tourist guides would try whenever possible to stand vis-à-vis in front of the person holding the map so that they could see (and read) the map in the right orientation, as shown on both images on Figure 5. A second tourist would often take a less active role beside them (e.g., not holding the map), thus forming a triangular F-formation. In this situation, although all three members have equal access to the physical space, their level of participation is unequal.

**Figure 6. Unusual F-formations in the wild. A couple adopts a back-to-back formation to provide protection from potential pickpocketing (left). A person taking a photo from the ground to capture the entire Duomo (right).**

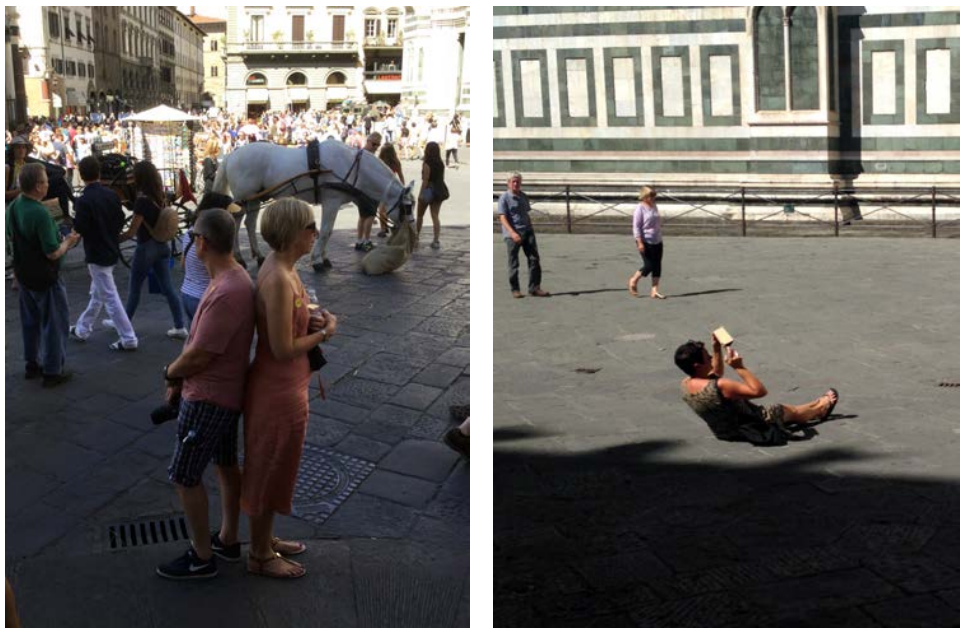

We also observed some unusual formations that have not been previously reported in F-formation studies conducted in controlled settings. In addition to the aforementioned resting on a bench back-to-back, lying flat, and in an inverted l-shape (Figure 2), we also saw a standing back-to-back formation (Figure 6, left). Taken in isolation, this F-formation seems almost unnatural. However, video analysis showed the man on the left repeatedly touching his back pocket, checking whether his wallet was still there. Moments before this picture was taken, the man and the woman were standing at a 90-degree angle with respect to each other. The man looked back at the woman and decided to press his body against hers in this back-to-back manner. Our interpretation is that standing back to back seemed to provide some sort of protection against potential pickpocketing. This image further helps illustrate the dynamic nature of F-formations in such contexts as this position was sustained for a mere 15 seconds. Another unusual situation consisted of a lady who broke away from a group to capture a picture of the Duomo. As she wanted to capture as much of the Duomo as possible, or perhaps from a particular angle, she decided to sit on the ground to take a low-angle shot (Figure 6, right). While this picture does not allow us to say much about F-formations themselves, it does again help us make a point about what we gained by going into the wild to make such observations, as these have not been reported in F-formation studies in controlled settings.

**Figure 7. Different uses of technology. A couple (left) and a girl in the foreground (right) using their smartphones while sitting side-by-side. A man in the background makes a Skype call using a headset (right).**

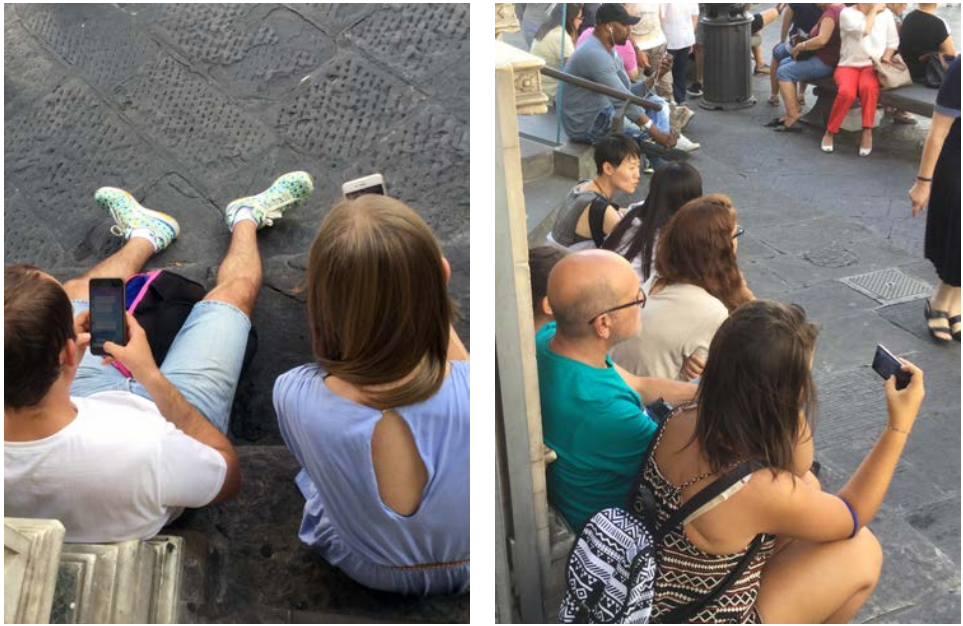

Finally, we also looked into people's use of their mobile devices and technology in general. Besides the aforementioned use of smartphones to take pictures and selfies, we observed several instances of people checking their email, reading text messages, and browsing the web while resting on benches or stairs side by side (Figure 7). Figure 7 shows a couple that are using a mobile device each (left), and two people sitting side-by-side where only the girl in the foreground is using her device (right). We also saw several individuals talking on their phones as they rushed from one place to another, or making a Skype call using a headset (Figure 7, right). People's use of their devices in dynamic larger-scale spaces seems to reflect how they naturally use technology in other more traditional environments (e.g., at home, work, or on a bus while commuting).

## DISCUSSION AND CONCLUSIONS

Beyond the specific F-formations watched during this exploratory observation, we can extract some lessons concerning the importance of in-the-wild studies and public observations for mobile collocation, namely the importance of the device ecology, the absence of directive task, and the freedom of social interactions.

We carried our observations in a public environment where people employed their own devices. Unlike "experimental" devices, which are given to participants for a specific study and thus are new to them, people use their personal devices in a very familiar and intimate way, which is often difficult to reproduce in the lab. The use of peripherals (such as headsets) and various types of phone case materials and shapes lead to particular ways using, holding, and interacting with their devices.

Contrary to a lab study, in our observations people were conducting their own tasks, such as wayfinding, usually intermixed with other activities, such as observing the environment, discussing with other people, or answering a phone call. This evolution between primary and secondary tasks has an impact on proxemics and on mobile interaction. For instance, F-formations change over time as people change their activities.

Finally, most of the people we observed seemed to be familiar with others around them. This led to unusual and very relaxed formations (such as when people were lying next to others on a bench). Obviously, these types of behaviors are very difficult to reproduce in experimental studies (whether they are in the lab or in the wild) where participants usually do not know each other.

The results of our exploratory F-formation observations described in this article are in contrast with the ones made in the literature on proxemics for mobile interaction. Our results highlight the importance of conducting not only in-the-wild studies, but also public observations, which can further inform the everyday usage of technology in collaborative situations.

# REFERENCES

Clawson, J., Voida, A., Patel, N., & Lyons, K. (2008). Mobiphos: A Collocated-Synchronous Mobile Photo Sharing Application. In *Proceedings of the 10th international conference on Human computer interaction with mobile devices and services* (MobileHCI '08). New York: ACM, 187-195. http://dx.doi.org/10.1145/1409240.1409261

Durrant, A., Rowland, D., Kirk, D.S., Benford, S., Fischer, J.E., & McAuley, D. (2011) Automics: souvenir generating photoware for theme parks. In *Proceedings of the SIGCHI Conference on Human Factors in Computing Systems* (CHI '11). New York: ACM, 1767–1776. http://doi.acm.org/10.1145/1978942.1979199

Greenberg, S., Hornbaek, K., Quigley, A., Reiterer, H., & Rädle, R. (2014). Proxemics in Human-Computer Interaction. *Dagstuhl Reports 3*, 11, 29–57. http://dx.doi.org/10.4230/DagRep.3.11.29

Greenberg, S., Marquardt, N., Ballendat, T., Diaz-Marino, R., & Wang, M. (2011). Proxemic Interactions: The New Ubicomp?. *interactions 18*, 1 (January 2011), 42–50. http://doi.acm.org/10.1145/1897239.1897250

Hall, E. (1963). A System for the Notation of Proxemic Behavior. *American anthropologist 65* 5 (October 1963), 1003–1026. http://dx.doi.org/10.1525/aa.1963.65.5.02a00020

Houben, S., & Marquardt, N. (2015). WatchConnect: A Toolkit for Prototyping Smartwatch-Centric Cross-Device Applications. In *Proceedings of the 33rd Annual ACM Conference on Human Factors in Computing Systems* (CHI '15). New York: ACM, 1247-1256. http://dx.doi.org/10.1145/2702123.2702215

Jokela, T., Chong, M.K., Lucero, A., & Gellersen, H. (2015). Connecting Devices for Collaborative Interactions. *interactions 22*, 4 (June 2015), 39-43. http://dx.doi.org/10.1145/2776887

Kendon, A. (1990). *Conducting Interaction: Patterns of Behavior in Focused Encounters*. Cambridge: Cambridge University Press.

Kortuem, G., Kray, C., & Gellersen, H. (2005). Sensing and Visualizing Spatial Relations of Mobile Devices. In *Proceedings of the 18th annual ACM symposium on user interface software and technology* (UIST '05). New York: ACM, 93–102. http://doi.acm.org/10.1145/1095034.1095049

Lucero, A., Clawson, J., Fischer, J., & Robinson, S. (2016a). Mobile Collocated Interactions With Wearables: Past, Present, and Future. *mUX J Mob User Exp 5*, 6. http://dx.doi.org/10.1186/s13678-016-0008-x

Lucero, A., Jones, M., Jokela, T., & Robinson, S. (2013). Mobile Collocated Interactions: Taking an Offline Break Together. *interactions 20*, 2 (March 2013), 26-32. http://doi.acm.org/10.1145/2427076.2427083

Lucero, A., Keränen, J., & Korhonen, H. (2010). Collaborative Use of Mobile Phones for Brainstorming. In *Proceedings of the 12th international conference on Human computer interaction with mobile devices and services* (MobileHCI '10). New York: ACM, 337-340. http://dx.doi.org/10.1145/1851600.1851659

Lucero, A., Quigley, A., Rekimoto, J., Roudaut, A., Porcheron, M., & Serrano, S. (2016b). Interaction Techniques for Mobile Collocation. In *Proceedings of the 18th International Conference on Human-Computer Interaction with Mobile Devices and Services Adjunct* (MobileHCI '16). New York: ACM, 1117-1120. http://doi.acm.org/10.1145/2957265.2962651

Marquardt, N., Hinckley, K., & Greenberg, S. (2012). Cross-device Interaction Via Micro-Mobility and F-formations. In *Proceedings of the 25th annual ACM symposium on User interface software and technology* (UIST '12). New York: ACM, 13-22. http://dx.doi.org/10.1145/2380116.2380121

Marshall, P., Rogers, Y., & Pantidi, N. (2011a). Using F-formations to Analyse Spatial Patterns of Interaction in Physical Environments. In *Proceedings of the ACM 2011 conference on Computer supported cooperative work* (CSCW '11). New York: ACM, 445-454. http://dx.doi.org/10.1145/1958824.1958893

Marshall, P., Morris, R., Rogers, Y., Kreitmayer, S., & Davies, M. (2011b). Rethinking 'Multi-User': an In-the-Wild Study of How Groups Approach a Walk-Up-and-Use Tabletop Interface. In *Proceedings of the SIGCHI Conference on Human Factors in Computing Systems* (CHI '11). New York: ACM, 3033-3042. http://dx.doi.org/10.1145/1978942.1979392

Mueller, F., Stellmach, S., Greenberg, S., Dippon, A., Boll, S., Garner, J., Khot, R., Naseem, A., & Altimira, D. (2014). Proxemics Play: Understanding Proxemics for Designing Digital Play Experiences. In *Proceedings of the 2014 conference on designing interactive systems* (DIS '14). New York: ACM, 533–542. http://doi.acm.org/10.1145/2598510.2598532

Paay, J., Kjeldskov, J., & Skov, M.B. (2015). Connecting in the Kitchen: An Empirical Study of Physical Interactions while Cooking Together at Home. In *Proceedings of the 18th ACM Conference on Computer Supported Cooperative Work & Social Computing* (CSCW '15). New York: ACM, 276-287. http://dx.doi.org/10.1145/2675133.2675194

Pearson, J., Robinson, S., & Jones, M. (2015) It's About Time: Smartwatches as public displays. In *Proceedings of the 33rd annual ACM conference on human factors in computing systems* (CHI '15). New York: ACM, 1257–1266. http://doi.acm.org/10.1145/2702123.2702247

Perrault, S.T., Lecolinet, E., Eagan, J., & Guiard, Y. (2013). Watchit: Simple Gestures and Eyes-Free Interaction for Wristwatches and Bracelets. In *Proceedings of the SIGCHI Conference on Human Factors in Computing Systems* (CHI '13). New York: ACM, 1451-1460. http://dx.doi.org/10.1145/2470654.2466192

Porcheron, M., Fischer, J.E., & Sharples, S. (2016). Using Mobile Phones in Pub Talk. In *Proceedings of the 19th ACM Conference on Computer-Supported Cooperative Work & Social Computing* (CSCW '16). New York: ACM, 1649-1661. http://dx.doi.org/10.1145/2818048.2820014

Santosa, S., & Wigdor, D. (2013). A Field Study of Multi-Device Workflows in Distributed Workspaces. In *Proceedings of the 2013 ACM international joint conference on Pervasive and ubiquitous computing* (UbiComp '13). New York: ACM, 63-72. http://dx.doi.org/10.1145/2493432.2493476

Serna, A., Tong, L., Tabard, A., Pageaud, S., & George, S. (2016). F-formations and collaboration dynamics study for designing mobile collocation. In *Proceedings of the 18th International Conference on Human-Computer Interaction with Mobile Devices and Services Adjunct* (MobileHCI '16). New York: ACM, 1138-1141. https://doi.org/10.1145/2957265.2962656

Terrenghi, L., Quigley, A., & Dix, A. (2009). A Taxonomy for and Analysis of Multi-Person-Display Ecosystems. *Personal Ubiquitous Comput. 13*, 8 (November 2009), 583-598. http://dx.doi.org/10.1007/s00779-009-0244-5

Tong, L., Serna, A., Pageaud, S., George, S., & Tabard, A. (2016). It's Not How You Stand, It's How You Move: F-formations and Collaboration Dynamics in a Mobile Learning Game. In *Proceedings of the 18th International Conference on Human-Computer Interaction with Mobile Devices and Services* (MobileHCI '16). New York: ACM, 318-329. http://dx.doi.org/10.1145/2935334.2935343